# Statistical evaluation and Phase Doppler Anemometry data processing of rotary atomization

Erika Rácz[a,*], Milan Malý[b], Ondřej Cejpek[b], Jan Jedelský[b], Viktor Józsa[a]

[a] Department of Energy Engineering, Faculty of Mechanical Engineering, Budapest University of Technology and Economics, Műegyetem rkp. 3., H-1111 Budapest, Hungary

[b] Faculty of Mechanical Engineering, Brno University of Technology, Technicka 2896/2, 616 69 Brno, Czech Republic

**Abstract**

Rotary atomization is used in a wide variety of fields, exploiting the external control option of the spray while no high-pressure fluid is needed. Most papers on rotary atomization deal with liquid jet breakup, while external spray characteristics are rarely evaluated; this is performed currently. The water spray was measured by a two-component Phase Doppler Anemometer. The optical setup requires a special measurement chamber to avoid spray deposition on the optical components. Therefore, the first goal was to find a proper filter that enables the removal of biased droplets by secondary flows. Since most droplets have a similar radial-to-tangential velocity ratio at each measurement point, i.e., scattering around a line, this was the first component of the best filter. The second component was the need for a positive radial velocity component. This filter efficiently removed droplets originating from alternative processes, increasing the $R^2$ of the line fit. The physical soundness of this filter was checked by evaluating the effect of filtering on the angle of the velocity components of each droplet at a given measurement point. The proposed filter efficiently detected recirculation, a secondary effect of the measurement setup with less regular data set shapes. Finally, the slope and intercept values of the fitted lines were evaluated and presented. The mean of the former followed the same trend irrespective of the rotational speed and the mass flow rate; it was principally dependent on the radial distance from the atomizer. The intercept showed a regular but less universal behavior.

Keywords: rotary atomizer, PDA, velocity components, regression, filter

**Nomenclature**

Latin letters

| Notation | Description | Unit (if relevant) |
|---|---|---|
| $A$ | slope of the fitted line | - |
| $B$ | intercept of the fitted line | m/s |
| $D$ | droplet diameter/size | μm |
| $f$ | filter | |
| $F$ | filtering ratio | % |
| $\dot{m}$ | mass flow rate | kg/h |
| min | minimum | |
| $n$ | rotational speed | rpm |
| $r$ | radial direction/axis | |
| $R^2$ | coefficient of determination | - |
| $SMD$ | Sauter mean diameter | μm |
| $t$ | tangential direction/axis | |
| $v$ | velocity | m/s |
| $z$ | axial direction/axis | |

Subscripts

| Notation | Description |
|---|---|
| f | filtered data |
| mag | magnitude |
| o | original data |
| $r$ | radial direction |
| $t$ | tangential direction |

Greek letters

| Notation | Description | Unit (if relevant) |
|---|---|---|
| $α$ | angle between the total velocity and radial direction | ° |
| $β$ | angle between the total velocity and tangential direction | ° |
| $Δ$ | change compared to the original data | |
| $σ$ | standard deviation | |

# 1. Introduction

Practical atomizers are mostly pressure or twin-fluid designs due to their simplicity over rotary atomizers, considering the family of mechanical atomizers. The critical advantage of rotary atomizers over the two other types is the ability to independently modify spray quality, i.e., droplet size distribution and the flow rate. Moreover, the flow field is only marginally affected by modifying the operating conditions [1]. Such atomizers are rarely used in current jet engines due to the moving parts and the hot environment, regardless of their fine spraying potential [2] and the resulting low $NO_X$ emission [3]. Nevertheless, its use and development for combustion applications several decades ago [4] notably contributed to our understanding of its fundamental working principles to employ this technology in alternative fields. Their use today has shifted to, e.g., the pharmaceutical industry [5], spray painting [6], food production [7], [8], wastewater treatment [9], non-invasive pipe repairing [10], and harvest aid sprayed from unmanned aerial vehicles [11], as a highlighted example among many in the agriculture. Rotary atomization is also efficient in metallurgy for processing slag, reducing waste [12], [13]. Moreover, it offers the opportunity to replace water quenching, enabling a water-free, efficient, and more flexible heat recovery method [14]. The ultimate advantage of rotary atomization for the industry is the wide and relatively uniform coverage over a short distance compared to other popular atomizer types [6], with easy control. Since most of the energy used to paint a car is related to the painting booth [15], process improvement directly offers significant savings on production costs [16]. Consequently, expanding the highly limited knowledge of rotary atomization available in the public literature, compared to pressure-swirl and twin-fluid atomizers, is timely and justified by the demand for numerous industrial applications.

The Sauter mean diameter, *SMD*, is the most important property of all evaporating sprays [1]. This parameter is governed by the geometry, rotational speed, *n*, and liquid properties [2].

Since droplets are formed from the ejected liquid jets via various breakup mechanisms, most studies in rotary atomization after the millennium focused on the disruption of these fluid structures, enabled by high-speed cameras with high quantum efficiency [17]. Obviously, the most influential parameter on jet breakup type is *n* [18], [19], which determines the jet thickness. This latter quantity marginally depends on the flow rate and strongly correlates with *SMD* [20], meaning that a size distribution can be shifted to the smaller droplets by enlarging the orifice diameter [21]. These findings ultimately imply the highly flexible operation in terms of turndown ratio.

Increased liquid viscosity leads to ligament formation from the jet that is still unstable. Therefore, a secondary breakup takes place, and finally, droplets are formed. Their *SMD* is increased compared to the utilization of less viscous liquids with quicker droplet formation from the jet since droplet formation occurs in a region with a reduced relative velocity between the gaseous and liquid phases [22]. Also, the droplet radial, $v_r$, and tangential velocities, $v_t$, decrease [23]. These velocity components are also critical in the spray structure of pressure-swirl atomizers [24]. The nearly constant value of jet thickness for a single liquid and *n* implies various breakup mechanisms as a function of the flow rate [25], as the jet velocity varies due to the law of continuity. The droplet size distribution may become multimodal since two or more dominant processes are possible [26]. Such behavior is undesired from the atomizer model development point of view since more physical parameters are needed for, e.g., an empirical *SMD* estimation formula. Consequently, such models quickly become highly inaccurate when extrapolation is required.

The typical layout of rotary atomizers is horizontal, while vertical alignment might cause uneven liquid distribution due to gravity. It can be balanced by increasing *n*; sufficiently uniform circumferential spray distribution can be achieved at about 30 m/s tangential velocities [27]. Since a fine spray is usually generated at high *n*, which comes with high tangential velocity, this is not an issue since most practical rotary atomizers operate at higher *n* [28].

Papers on external spray characteristics of rotary atomizers are mostly confined to *SMD* and, less frequently, droplet size distribution. However, knowing the velocity field is essential for designers to find the best atomizer geometry and *n* for a given application [1]. The phase Doppler technique is the best tool for non-invasive size and velocity measurement [29], which is also employed in the present study. The jet velocity at the orifice outlets equals $v_t$ and is in the range of 10 m/s downstream [23]. This represents significant momentum, justifying the need for a comprehensive flow field characterization and a two-component Phase Doppler Anemometer, PDA, for optical spray measurement. The third velocity component, the axial movement of droplets, is usually minor compared to the other two components but characterizes the spray cone angle, which is critical for, e.g., spray cooling applications [30].

Measurements with PDA produce a large amount of data, which should be filtered to remove artefacts [31]. However, some data sets may show a biased physical behavior [32], which is challenging to address if the causing phenomenon is not entirely explored. Since the nature of the noise could vary between physical fields, there is no universal approach to address them. Therefore, the present paper evaluates various simple filters to keep the process robust and less affected by the data set properties. The corresponding details are discussed in Subsection 2.2.

The novelty of this paper is the following. It is known from the operation principle of rotary atomizers that the radial and tangential velocity components are both important along the spray. At the orifice, the tangential velocity of the liquid jet equals that of the solid body, which does not hold downstream. Also, the velocity decay of a potential vortex cannot be used for flow field characterization since the induced flow field of the upcoming jets pushes the spray further away from the atomizer. The accompanying problem of PDA measurement is proper data filtering since the complex flow structures might trap droplets in vortices, which were injected long ago and originated from the required optical setup of such measurements. This problem is addressed to reduce the measurement bias of velocity field analysis, which is a critical step toward spray model development and numerical model validation of rotary atomizers.

## 2. Materials and methods

The experimental setup is detailed in Subsection 2.1, detailing the key parameters of the optical setup, the atomizer, and the measurement conditions. Subsection 2.2 focuses on raw PDA data processing of rotary atomization in general, highlighting the non-trivial filtering procedure.

### *2.1 Experimental setup*

The rotary atomizer with a 110 mm disk outer diameter was investigated; see the schematic drawing of the atomizer test bench and the PDA system in Fig. 1. The spray was generated from two orifice sets with four orifices each. The first set had an orifice diameter of 1.2 mm and was located in the center of the PDA measurement at $z = 0$. The diameter in the case of the second orifice set was 1.5 mm and was axially shifted by $z = 4.5$ mm. This design aims to investigate the effect of orifice diameter in a single measurement via spray symmetry and to discuss possible spray interactions via statistical methods. The rotary atomizer was operated in $n = 5{,}000\text{–}30{,}000$ rpm (83 to 500 Hz). Spray collection was performed in a chamber with a 400 mm diameter. The optical access was provided via the partially open geometry from both the front and rear sides. This solution was necessary since the spray droplets with high radial velocity would quickly cover any window, leading to biased results. A purging air curtain was also unsuitable due to its significant flow field-altering effect. Demineralized water at 22 °C temperature was used as the test liquid. A frequency-driven centrifugal pump fed it to the atomizer through a Corioliss Optimass 6000 mass flow meter (Krohne, Germany). The liquid was injected using six tubes to form a homogeneous liquid film on the internal surface of the rotating disk. The mass flow rate, $\dot{m}$, varied from 36 to 288 kg/h (10 to 80 g/s) with < 0.2% uncertainty of the measured value.

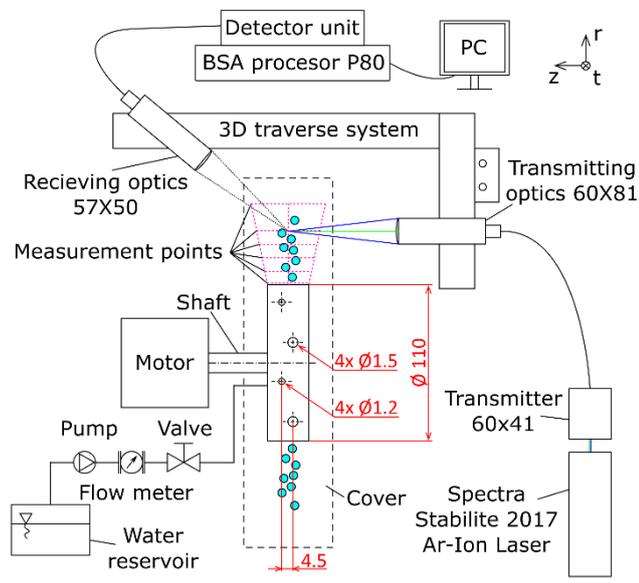

**Figure 1.** Schematic drawing of the rotating atomizer and PDA system. Not-to-scale with dimensions in mm.

The size and velocity of the spray droplets were measured with a two-component fiber-based commercial PDA (Dantec Dynamics A/S Skovlunde, DK). This system acquired the tangential and radial velocity components in coincidence mode along with the simultaneous drop sizes in the radial direction. The spray was probed at radial distances from the rotary atomizer of $r$ = 5, 10, 15, 20, and 30 mm. An axial line was measured on each radial distance with 13 to 21 positions on each axis, following the axial spreading of the spray with increasing radial distance. The step size on each axial line was uniformly 2 mm. Each measurement point endured 7 s or collecting 35,000 spherical droplet samples, whichever condition was fulfilled earlier, resulting in fewer samples towards the periphery. The PDA was aligned using a scattering angle of 41° to allow the measurement inside the collecting chamber. The focal length was 500 mm for both the transmitter and the receiver. The receiver was equipped with mask B and a spatial filter width of 0.1 mm. The formed measuring volume was 0.12×0.12×0.16 mm in the radial, tangential, and axial directions, respectively, and the optical setup allows a maximum droplet size measurement of 131 μm. Further details of the measurement setup can be found in [33].

The range of the measurement velocity was set from 0 to 154 m/s for the tangential ($t$) velocity component and from -18 to 128 m/s for the radial ($r$) velocity component. The measurement points are presented in Fig. 2, detailing the area in Fig. 1 with magenta dots. Considering the five $\dot{m}$, six different $n$ values and 85 positions in the spray, the total number of evaluated measurement points is 2,550. All the measurement parameters are included in Table I.

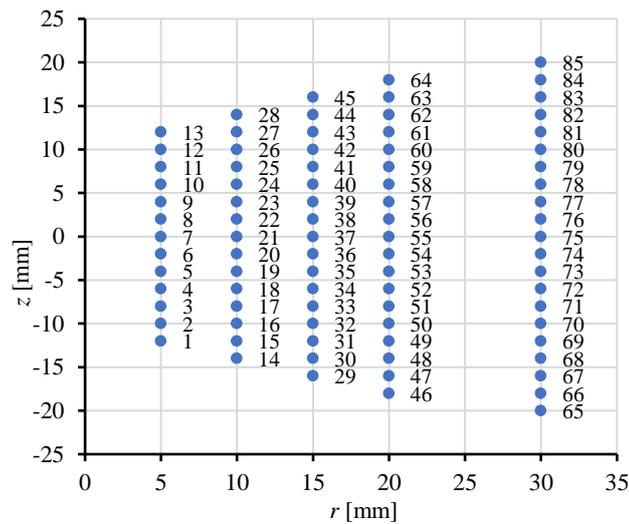

**Figure 2.** Measurement points measured from the disk surface and along the axis.

**Table I.** Measurement points and settings.

| $r$ [mm] | 5 | 10 | 15 | 20 | 30 |
|---|---|---|---|---|---|
| $z$ [mm] | -12–12 | -14–14 | -16–16 | -18–18 | -20–20 |
| $n$ [krpm] | 5, 10, 15, 20, 25, 30 ||||| 
| $\dot{m}$ [g/s] | 10, 20, 40, 60, 80 |||||

*2.2 Data processing*

The raw PDA data sets were first filtered with a high-pass 0.5145 μm filter to keep droplets exceeding the laser wavelength in size. Since the droplets generated by a rotary atomizer possess significant radial and tangential velocity components ($v_r$, $v_t$), there are several possibilities for data filtering to remove the non-representative droplets, which might be trapped in a vortex and not part of the main flow field. This is partially due to a bias caused by the physical setup to keep the droplets away from the optical setup. Since most data sets showed a scatter around an inclined line with various slopes in $v_r$–$v_t$ plots, the ultimate evaluation criterion was the overall improvement in the coefficient of determination, $R^2$, of the fitted line considering all data sets. If non-characteristic data points are efficiently removed, the fit quality increases. However, the number of removed data points was also checked to avoid using excessively aggressive filtering methods.

Among the various filters, the *isoutlier* Matlab function was used, which determines the median absolute deviation and filters data points with three times the median absolute deviation or further away from the median. This filtering function performed excellently for both pressure and twin-fluid atomizers in a preceding work due to its robustness for non-Gaussian probability distributions compared to other methods such as IQR or Z-score. [34]. Machine learning methods cannot classify data universally at each measurement point due to the changing multimodal, correlated bivariate characteristics of both droplet size distributions and distribution of velocity components. Therefore, a universal filtering method is needed to choose a suitable variable that can handle the changing distribution of the data and the outliers in terms of both radial and tangential velocities.

The following developed filters were all performed with the introduced *isoutlier* function but different variables were defined to be investigated for getting the desired result matching the underlying physical phenomena. The filters below are marked with *f* with a subscript with the quantity for filtering. It can be used for each of the measured quantities of the PDA: $v_r$, $v_t$, and droplet size, $D$, which was determined in the radial direction. These are filters 1-3, $f_{v_r}$, $f_{v_t}$, and $f_D$. The fourth one is a double filter for outliers in either $v_r$ or $v_t$, $f_{v_r \vee v_t}$, where ∨ is the logical 'or' operator. Since $v_r$ and $v_t$, both dominate the droplet velocity, a fifth filtering method, $f_{v_{mag}}$, can also be performed on the velocity magnitude, $v_{mag}$, estimated as:

$$v_{mag} \approx \sqrt{v_r^2 + v_t^2}. \tag{1}$$

The following filters are more advanced and can only be recommended for rotary atomizers, while the first five are general and can be effectively used for other atomizer types. The sixth one is exploiting the fact that droplets should possess a positive radial velocity, $f_{v_r > 0}$. Since the jet exits the orifice at a significant radial velocity that is decreased over time, negative velocity most probably characterizes droplets trapped in a vortex. Since the droplets scatter around a line, filtering by the velocity ratio, $v_r/v_t$, $f_{v_r/v_t}$ is also meaningful, being the seventh filter. Since this ratio more aggressively removes droplets with relatively low $v_t$ values due to the hyperbolic function and is more friendly with outlier $v_r$ values, the reciprocal of this ratio, $v_t/v_r$, $f_{v_t/v_r}$ is the eighth filtering approach. The ninth method was a combination of the sixth and seventh filters, i.e., outliers in either $v_r/v_t$ or $v_r > 0$, $f_{v_r/v_t \vee v_r > 0}$. The last method $f_{v_r > 0, v_r/v_t}$ was similar to $f_{v_r/v_t \vee v_r > 0}$, but first, the $v_r > 0$ condition and then outlier filtering by $v_r/v_t$ was applied. Methods in connection with size-velocity correlation are not involved due to the $f_D$ filters non-compliance, which is detailed in the beginning of Appendix A. Other combinations were also tested without notable improvement compared to the last two methods. The corresponding results are discussed in Subsection 3.1. The steps of the analysis is explained in Fig. 3.

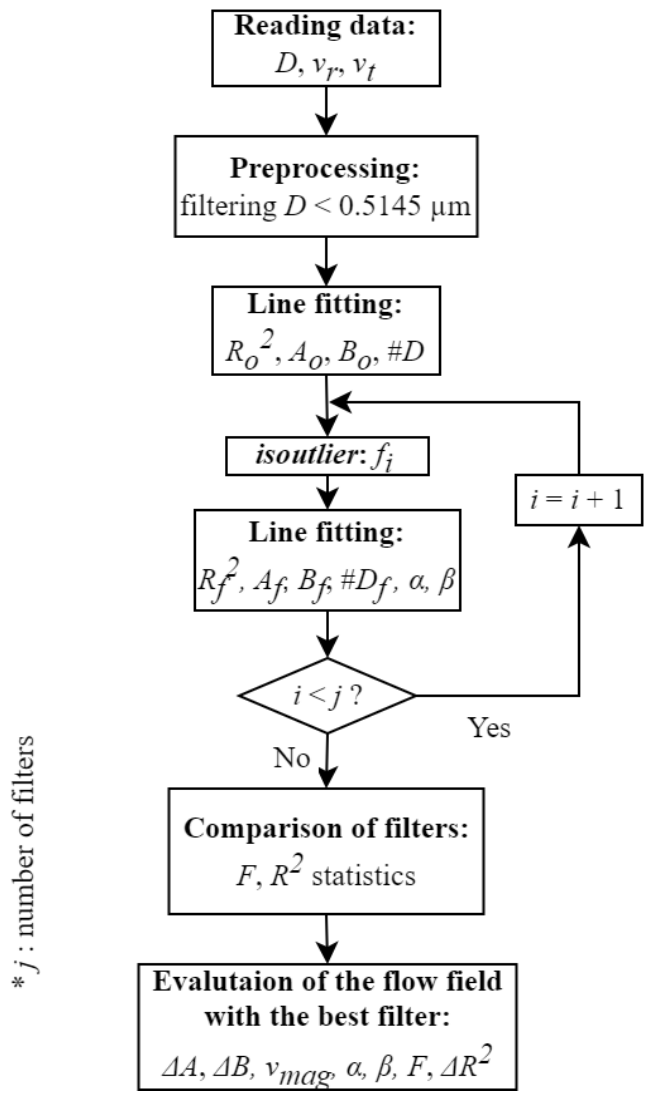

**Figure 3.** Steps of data processing.

## 3. Results and discussion

Firstly, the trends of the $v_r$–$v_t$ plots and the line fitting are introduced at different measurement settings. The effect of the $f_{v_r/v_t \vee v_r > 0}$ filter is shown in Subsection 3.1, while other filters are detailed in the Appendix. The second subsection shows the effect of the recirculation on the flow field and the filtering results. Finally, Subsection 3.3 shows the results of the line fitting in terms of measurement settings.

*3.1 Evaluating the filtering methods*

The $v_t$ and $v_r$ of each droplet in rotary atomization scatter around an inclined line with various slopes, depending on the atomization parameters and the position of the measurement point in the spray. A few examples are shown in Fig. 4., highlighting four plots; each droplet is colored by its diameter. These plots represent points close and far from the nozzle, at the center, and close to the periphery. The range of $v_r$ is similar for all the positions in the spray, but $D$ and $v_t$ differ significantly. The linear relationship between the two velocity components is more apparent close to the nozzle, while the slope decreases towards the periphery and downstream the nozzle. Since larger droplets lose their momentum slower than smaller ones due to the aerodynamic drag, most droplets with $v_r$ and $v_t$ velocities are relatively large. All plots are polluted by droplets of various $D$ with low $v_t$ but high $v_r$. These make any model fitting process conclude with a notable bias since their number is low, but their distance from the bulk data is significant. Therefore, filtering is essential to end up with more representative spray data.

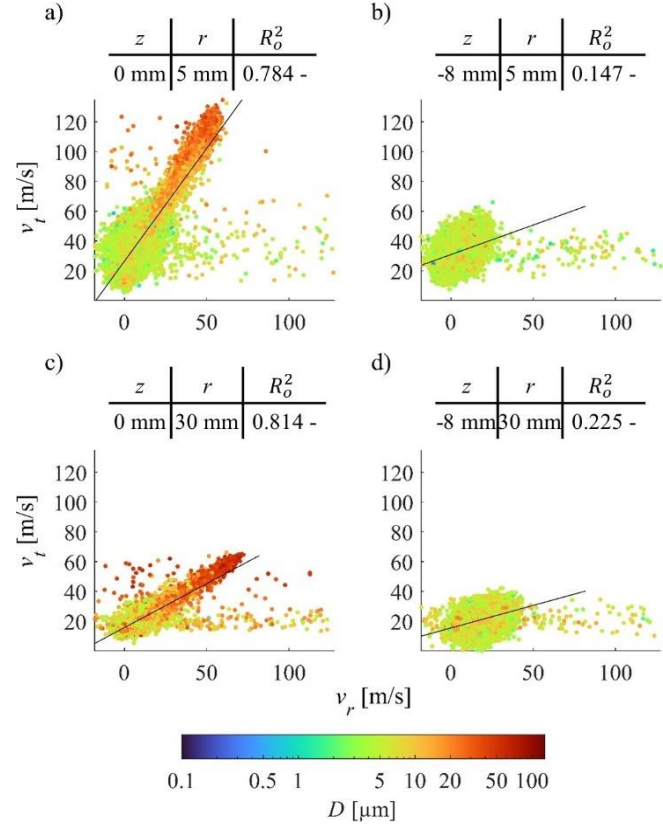

**Figure 4.** Radial and tangential velocity distribution of the spray, colored by the droplet diameter at $n$ = 30 krpm and $\dot{m}$ = 10 g/s. $R_o^2$ refers to the coefficient of determination of the fitted line in the case of the original data.

Depending on $n$ and $\dot{m}$, the distribution of the points may change on the $v_r$–$v_t$ plane, for example, in Fig. 4. At $z$ = 0 mm, the distribution is more elongated, while, at the periphery, it is closer to an ellipse without size-velocity correlation. Droplets originating from breakup phenomena with low probability under the given conditions may result in alternative linear trends, e.g., a horizontal line at $v_t$ = 20 m/s in Fig. 4c. Outliers appear in the data with various frequencies and locations depending on the measurement settings.

The filtering aims to remove data that differs significantly from the linear trend in the radial-tangential velocity field by choosing the suitable variable as criteria. The filters detailed in Subsection 2.2 are used on all data, and their effect on the line fit was evaluated in terms of coefficient of determination, $R^2$. The results of the line fit without any filter are shown in the first row of Table II., which is the reference. The mean $R^2$ of filters 1-6 were found to be lower compared to the original data sets. However, the standard deviation, $\sigma$, shows a slight reduction in the first five cases, meaning more consistent data. The minimum $R^2$ values are similarly low in each case, while the maximum $R^2$ value is slightly smaller in the case of filters 1–6 compared to case 0, the original data. The effects of these filters are shown in the Appendix in Figs. A.1–6. The last column contains the cardinality of the remaining droplets after filtering, $\#D_f$, compared to the total number of droplets, $\#D$. The result is called the filtering ratio, $F$, calculated as:

$$F = \frac{\#D_f}{\#D} \cdot 100\%. \tag{2}$$

**Table II.** The effect of various filters. Mean of $R^2$, the standard deviation of $R^2$, the minimum of $R^2$, the maximum of $R^2$, and the mean of $F$. Boldface denotes the four best-performing filters. Since 9 and 10 are very close to each other, filters 7–9 are presented further to assess their performance.

| No. | $f$ | $\overline{R^2}$ | $\sigma_{R^2}$ | $\min(R^2)$ | $\max(R^2)$ | $\overline{F}$ [%] |
|---|---|---|---|---|---|---|
| 0 | original data | 0.4046 | 0.2858 | -0.0009 | 0.9602 | - |
| 1 | $D$ | 0.3029 | 0.2584 | -0.0009 | 0.9465 | 7.923 |
| 2 | $v_t$ | 0.258 | 0.2358 | -0.0009 | 0.9477 | 4.935 |
| 3 | $v_r$ | 0.3314 | 0.26 | -0.0011 | 0.9584 | 4.911 |
| 4 | $v_{mag}$ | 0.2912 | 0.2421 | -0.0009 | 0.9491 | 5.947 |
| 5 | $v_r \lor v_t$ | 0.3144 | 0.2578 | -0.0011 | 0.9576 | 6.123 |
| 6 | $v_r > 0$ | 0.3979 | 0.2934 | -0.0018 | 0.9562 | 25.54 |
| 7 | $v_r/v_t$ | **0.4682** | **0.3083** | **-0.0007** | **0.9734** | **2.356** |
| 8 | $v_t/v_r$ | **0.4134** | **0.2893** | **-0.001** | **0.9624** | **16.23** |
| 9 | $v_r/v_t \lor v_r > 0$ | **0.4938** | **0.3106** | **-0.00055** | **0.9821** | **27.13** |
| 10 | $v_r > 0, v_r/v_t$ | **0.4938** | **0.3106** | **-0.00055** | **0.9821** | **27.09** |

The superior filters consider the ratio of the two velocity components; the $R^2$ increases in cases 7–10. The seventh filter, $f_{v_r/v_t}$, shows relatively high improvement in $R^2$ with low $F$ and being physically meaningful, as shown in Fig. A.7. In the case of $f_{v_t/v_r}$, in contrast with the improvement in $R^2$, the results are incorrect in terms of the physical meaning, detailed further in the Appendix.

The most significant improvement was achieved in the case of $f_{v_r/v_t \vee v_r > 0}$ and $f_{v_r > 0 + v_r/v_t}$. $F$ is the highest in these two cases due to the numerous negative radial velocity droplets. The statistics of $R^2$ are very similar in both cases; therefore, the line fitting is not sensitive to the order of filtering. Only $F$ shows a 0.04% difference, which means an average of 14 droplets as of a 35.000 droplet dataset.

**Table III.** Improvement in $R^2$ of various filters. Mean, standard deviation, minimum, and maximum. The boldface denotes the four best-performing filters. Since 9 and 10 are very close to each other, filters 7–9 are presented further to assess their performance.

| No. | $f$ | $\overline{\Delta R^2}$ | $\sigma(\Delta R^2)$ | $\min(\Delta R^2)$ | $\max(\Delta R^2)$ |
|---|---|---|---|---|---|
| 1 | $D$ | -0.1017 | 0.1277 | -0.7139 | 0.1918 |
| 2 | $v_t$ | -0.1466 | 0.1814 | -0.859 | 0.1411 |
| 3 | $v_r$ | -0.0731 | 0.1866 | -0.8242 | 0.5538 |
| 4 | $v_{mag}$ | -0.1134 | 0.194 | -0.824 | 0.552 |
| 5 | $v_r \vee v_t$ | -0.0902 | 0.1885 | -0.8223 | 0.554 |
| 6 | $v_r > 0$ | -0.0067 | 0.0318 | -0.1861 | 0.2629 |
| 7 | $v_r/v_t$ | **0.0637** | **0.1419** | **-0.7221** | **0.6413** |
| 8 | $v_t/v_r$ | 0.0088 | 0.0344 | -0.26 | 0.5899 |
| 9 | $v_r/v_t \vee v_r > 0$ | **0.0892** | **0.1454** | **-0.7036** | **0.7597** |
| 10 | $v_r > 0 + v_r/v_t$ | **0.0892** | **0.1415** | **-0.7036** | **0.7597** |

The relative indicators are shown in Table III to quantify the performance of the filters compared to case 0. Filters 7–10 only improve $R^2$, which was already presented in Table II. Since it is desired to positively impact the fit quality, the first six filters can be immediately discarded. Excessively positive discrimination would otherwise lead to critical information loss, meaning that the goal is a physically sound filter, not the maximization of $R^2$. Filters 6 and 8 are neutral, indicated by closely zero $\overline{\Delta R^2}$ and $\sigma(\Delta R^2)$ values, which also confirms the non-compliance of $f_{v_t/v_r}$ and $f_{v_r > 0}$. Consequently, removing the negative velocity terms does not lead to significant improvement, regardless of the more than 25% data removal. However, it is physically justified.

Considering that outliers need to be removed, improved fit quality is expected in the particular case of rotary atomization, where the linear velocity ratio is present. Therefore, the remaining options include $v_r/v_t$; the difference is the further consideration of the implementation of combining $v_r/v_t$ and $v_r > 0$. The improvement with this latter condition is apparent; the difference is the further removed data points, meaning a more significant data cut. The last two columns indicate the worst and the best effects of the filters on $R^2$. The smallest negative effect, such as the highest positive effect, was in the case of filters 9 and 10.

$f_{v_r > 0 + v_r/v_t}$ seems to have a more logical condition order for the filtering. However, $f_{v_r/v_t \lor v_r > 0}$ is accepted as the best solution for the problem since no notable improvement was achieved by prefiltering the data sets for $v_r > 0$. In the recirculation zones, most droplets have negative radial velocities. However, some might have a positive radial velocity tending towards the negative velocity range due to the direction of the flow. The multimodal distributions of the radial velocity occur due to the combination of different atomization processes and recirculation, and besides, their characteristics cannot be separated. Some droplets may appear as outliers by $v_r/v_t$ and have negative radial velocities but do not differ significantly from the bulk data. Therefore, the joint conditions of the $f_{v_r/v_t \lor v_r > 0}$ filter seems to be the proper method.

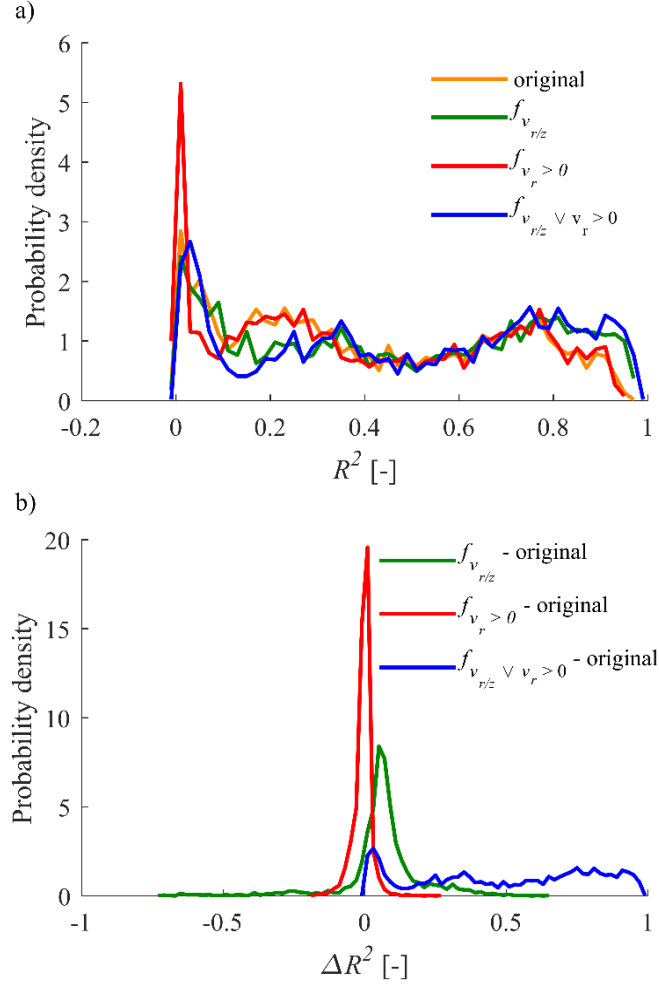

**Figure 5.** a) Histogram of $R^2$ in the original and the $f_{v_r/v_t \vee v_r > 0}$ case. b) Histogram of the improvement in $R^2$.

The beneficial effect of $f_{v_r/v_t \vee v_r > 0}$ is apparent in the less regular data sets, where the linear model failed, and in regions where such a model worked already well. The effect of filtering on $R^2$ is shown in Fig. 5; the mean value of $\Delta R^2$ is 0.0892, shown in Table II. The extreme effects on data sets are not typical; mostly, a slight improvement is made, which was desired. The $f_{v_r > 0}$ alone does not provide sufficient data removal since the high-velocity, non-representative droplets are kept. However, $f_{v_r/v_t}$ alone performs well, being the best single-step filter. Based on the superiority of $f_{v_r/v_t \vee v_r > 0}$, if filtered data are mentioned in the remaining parts of the paper, it refers to this specific filter, and filtered data sets get an *f* subscript, while *o* subscript refers to the original, unfiltered data.

Figure 6 shows the effect of the $f_{v_r/v_t \lor v_r > 0}$ filter on the dataset of Fig. 4. The red line is the fitted line of the filtered data set, which shows improvement in terms of $R^2$ in all the presented cases, which is often accompanied by an increase in slope. The filtered droplets are marked with a black circle. The $f_{v_r/v_t \lor v_r > 0}$ filter usually removes more droplets close to the nozzle, where *F* is above 20%, while far from the nozzle, *F* is generally below 10%. At $r = 30$ mm, for example, Figs. 6c and 6d, the slope and the intercept of the fitted line do not change remarkably, while the improvement of $R^2$ is 0.08–0.09. In the centerline at $r = 5$ and 30 mm, Figs. 6a and 6c, the filtering removes most outliers far from the fitted line. At the periphery of the spray, such as Figs. 6b and 6d, there is a visible difference between the shape of the points distribution and the fitted line. Close to the nozzle, in Fig. 6b, the fitted line does not follow the shape of the remaining droplets after the filtering, and the improvement in $R^2$ is minimal. Far from the nozzle, in Fig. 6d, the filtering improves fitting and $R^2$.

In some cases, droplets with high tangential and low radial velocity appear without filtering, such as in Figs. 6a and 6c. These droplets are relatively big; therefore, their tangential velocity is higher due to their inertia. Their distribution shows no trend. Consequently, removing these droplets is irrelevant during the outlined problem. They appear with significantly lower frequency than the outliers with high radial and low tangential velocities. From these results, it can be concluded that filtering works properly on all data except the vicinity of the nozzle at low *n*.

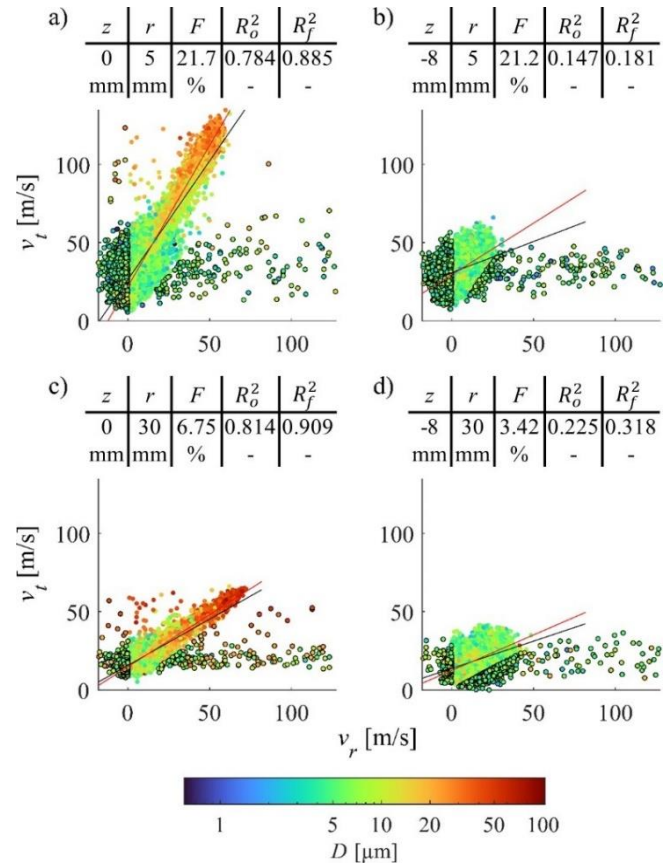

**Figure 6.** Radial and tangential velocity distribution of the spray, colored by the droplet diameter at $n$ = 30 krpm and $\dot{m}$ = 10 g/s, using the $f_{v_r/v_t \vee v_r > 0}$ filter. The black line is fitted to the original data set, while the red line is fitted to the filtered data set; the discarded droplets are marked with black circles.

The effect of $\dot{m}$ and $n$ is presented in Fig. 7, using the $f_{v_r/v_t \vee v_r > 0}$ filter. The first column shows the velocity distributions at $r$ = 5 mm and $z$ = 0 mm, and the second row shows them close to the periphery, $r$ = 5 mm and $z$ = -8 mm. The data sets of further radial distances show very similar behavior, which are not presented here.

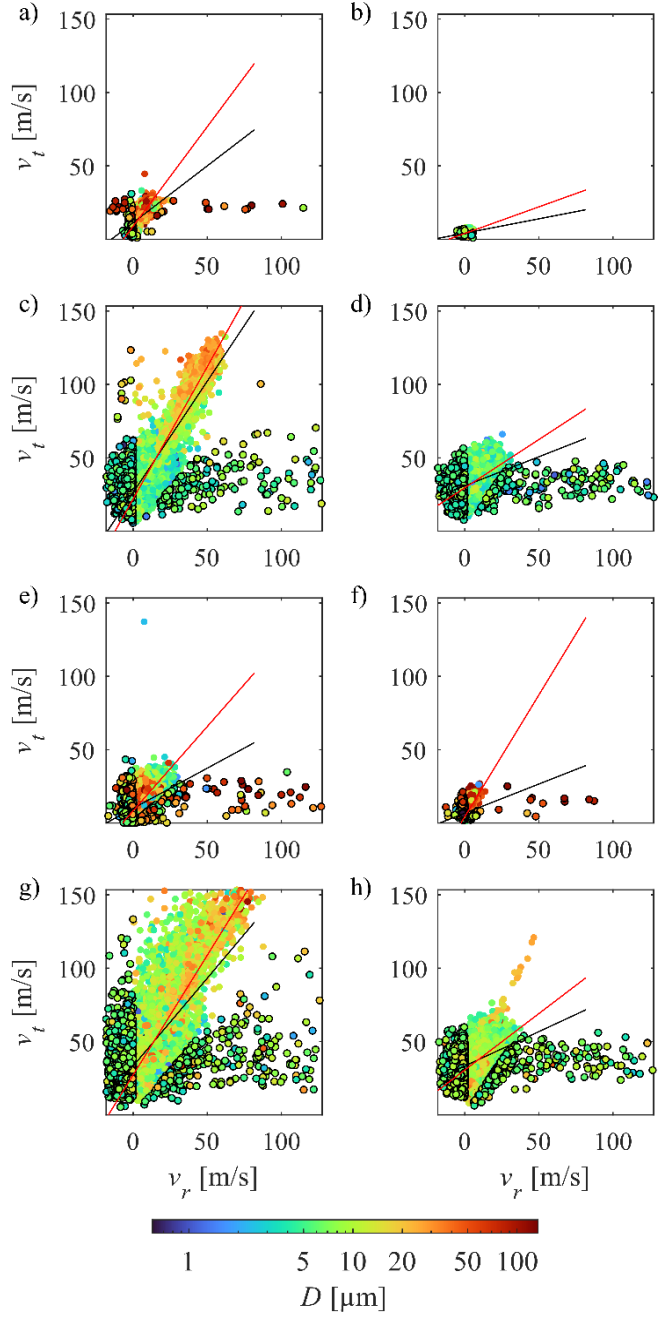

**Figure 7.** Radial and tangential velocity distribution of the spray, using the $f_{v_r/v_t \lor v_r > 0}$ filter, colored by the droplet diameter. The black line is fitted to the original data set, while the red line is fitted to the filtered data set; the discarded droplets are marked with black circles. Table IV. contains the measurement settings of each figure.

**Table IV.** Measurement setting of Figure 7a–h.

| No. | $z$ [mm] | $r$ [mm] | $\dot{m}$ [g/s] | $n$ [krpm] | $\overline{R_o^2}$ [-] | $\overline{R_f^2}$ [-] | $F$ [%] |
|---|---|---|---|---|---|---|---|
| a) | 0 | 5 | 10 | 5 | 0.444 | 0.652 | 26.7 |
| b) | -8 | 5 | 10 | 5 | 0.08 | 0.144 | 40.1 |
| c) | 0 | 5 | 10 | 30 | 0.784 | 0.885 | 21.8 |
| d) | -8 | 5 | 10 | 30 | 0.147 | 0.181 | 21.2 |
| e) | 0 | 5 | 80 | 5 | 0.287 | 0.514 | 38.4 |
| f) | -8 | 5 | 80 | 5 | 0.158 | 0.491 | 35.4 |
| g) | 0 | 5 | 80 | 30 | 0.52 | 0.666 | 19.35 |
| h) | -8 | 5 | 80 | 30 | 0.163 | 0.249 | 9.02 |

In Figs. 7a, 7b, 7e, and 7f, at low *n*, the droplets are scattering in a small circular area close to the origin, especially at the periphery of the spray. However, a significant increase can be achieved in $R^2$ by using $f_{v_r/v_t \vee v_r > 0}$. In such cases, higher *F* and a notable change in the slope and the intercept of the fitted line are also observed. At higher frequency and close to the centerline, such as in Figs. 7c and 7g, where elliptical droplet scatter is observable, higher $R^2$ is achieved with lower *F*. In the case of Figs. 7d and 7h, at the periphery of the spray, recirculation and air entrainment effects are enhanced. It results in a similar distribution of the droplets to the case of Figs. 7a, 7b, 7e, and 6f but with higher negative and outlying radial velocities. The $f_{v_r/v_t \vee v_r > 0}$ results in a limited improvement in $R^2$ but a noticeable change in the slope and intercept of the fitted line. In Fig. 7h, the appearance of the droplets with high radial and tangential velocity follows the expected linear characteristics, indicating less appropriate line fitting, even though $R^2$ improves.

*3.2 Flow direction and recirculation*

Due to the force exerted by the rotating drum, the droplets follow a ballistic pathway, depending on the measurement settings. The angle between the droplet velocity components, *α*, is calculated as:

$$\alpha = \arctan\left(\frac{v_t}{v_r}\right). \qquad (3)$$

To visualize the same data in other view, the complementary angle of *α* is calculated:

$$\beta = \begin{cases} 90° - \alpha, \text{if } \alpha > 0°; \\ -90° - \alpha, \text{if } \alpha < 0°. \end{cases} \qquad (4)$$

The sign of $\alpha$ only depends on the sign of $v_r$, because $v_t$ is always positive. Therefore, droplets trapped in a recirculation zone can be distinguished by their negative $v_r$, and they point towards the nozzle.

$\alpha$ of each droplet as a function of $v_r$ is shown in Fig. 8 at the same measurement settings as in Fig. 6, highlighting the filtered data. In the central region of the spray, e.g., in Figs. 8a and 8c, crossing hyperboles appear, which belong to different atomization processes, but both are described with linear $v_r$–$v_t$ relation, shown in Figs. 6a and 6c. As the result of the filtering, in the central part of the spray, $\alpha$ scattered around a hyperbole. Larger droplets usually have higher $v_r$ and lower $\alpha$. In Figs. 8b and 8d, the filtered droplets with positive radial angles match the trend of the hyperbole of the unfiltered droplets. However, if these are compared with Figs. 6b and 6d, these droplets differ from the typical velocity distribution observed at the spray periphery. The location of the unfiltered droplets around a hyperbole is wider due to the non-elongated $v_t$–$v_r$ distribution, and there is no tendency between $D$ and the $\alpha$ or $v_t$.

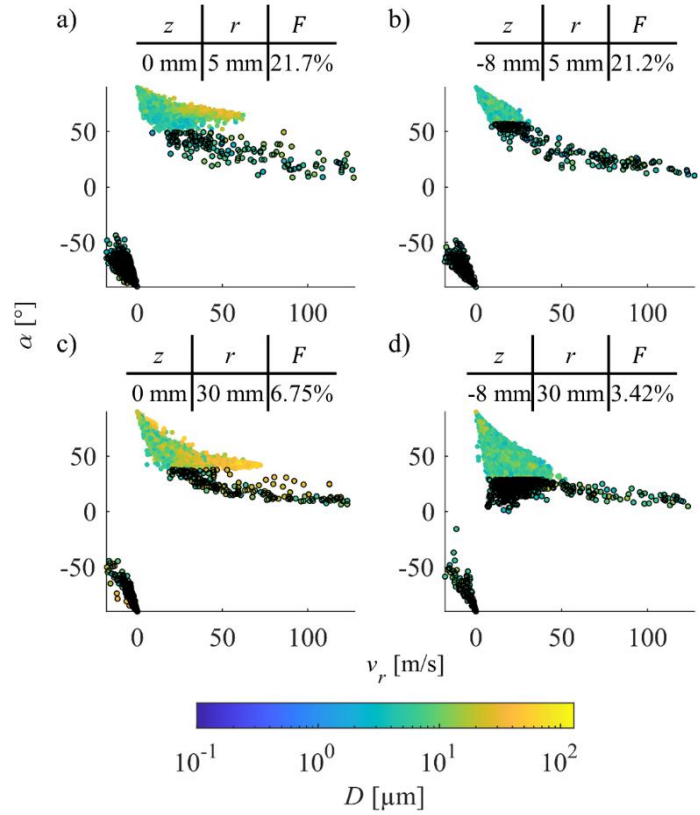

**Figure 8.** Relation of $v_r$ and $\alpha$ of the unfiltered data, colored by the droplet diameter at $n$ = 30 krpm and $\dot{m}$ = 10 g/s, using the $f_{v_r/v_t \vee v_r > 0}$ filter. The discarded droplets are marked with black circles.

Figure 9 shows $\beta$ as a function of $v_t$ at identical measurement settings to Fig. 8. Due to the negative $\alpha$, droplets with $\beta < 0°$ are filtered. The other filtered droplets with positive $\beta$ are removed because they originate from undesired physical processes. Closer to the nozzle, in Fig. 9a, the shape of the distribution is more elongated due to the large droplets moving by high inertia. In Fig. 9c, the flow direction of larger droplets results in higher $\beta$ values.

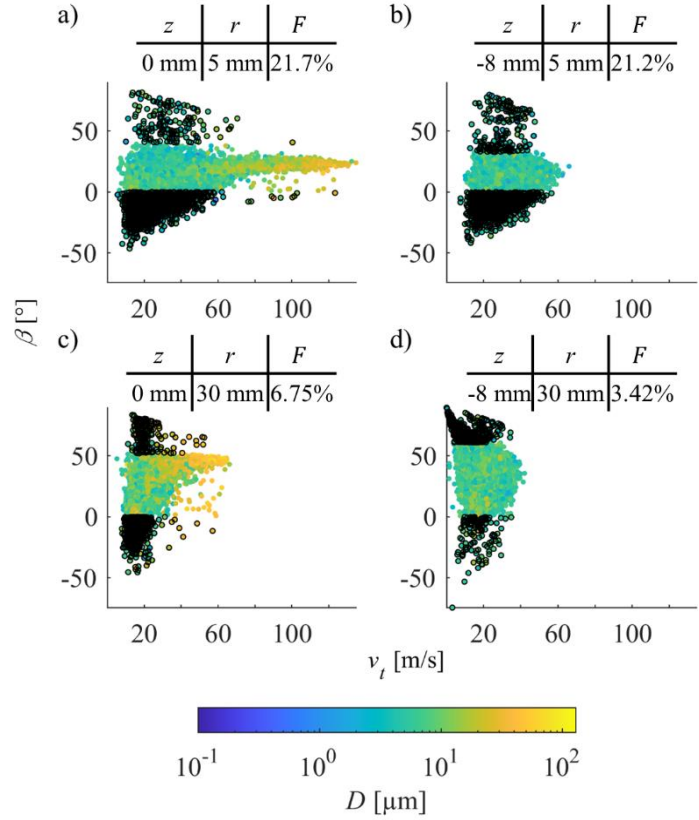

**Figure 9.** Relation of $v_t$ and $\beta$ of the unfiltered data, colored by the droplet diameter at $n = 30$ krpm and $\dot{m} = 10$ g/s, using the $f_{v_r/v_t \vee v_r > 0}$ filter. The discarded droplets are marked with black circles.

The $v_t$ is determined only by $n$, while $v_r$ is influenced by multiple processes, such as aerodynamic drag and turbulent dispersion. Therefore, $v_r$ depends on $v_t$, but $v_t$ is not dependent on $v_r$. As a result, $\beta$ is nearly constant as a function of $v_t$, while $\alpha$ shows a hyperbolic trend as a function of $v_r$.

The radial angles are calculated from the arctangent of the ratio of the velocity components. Therefore, the outlier filtering results in a cut of $\alpha$ and $\beta$ distributions. $\alpha$ or the hyperbole of the remaining droplets converges to a value based on the spray parameters. In the case of all the investigated spray parameters, Fig. 10 shows the minima of the radial angles as a function of frequency, mass flow rate, and measurement point. Based on the results, the minimum radial angle does not depend on the mass flow, unlike frequency and the measurement point. At $z > 0$ and the periphery of the spray, the minimum angle increases drastically due to the recirculation, except at $n = 5$ krpm due to recirculation. Hence, this indicator clearly shows the zone where the measurement setup significantly biased the flow field.

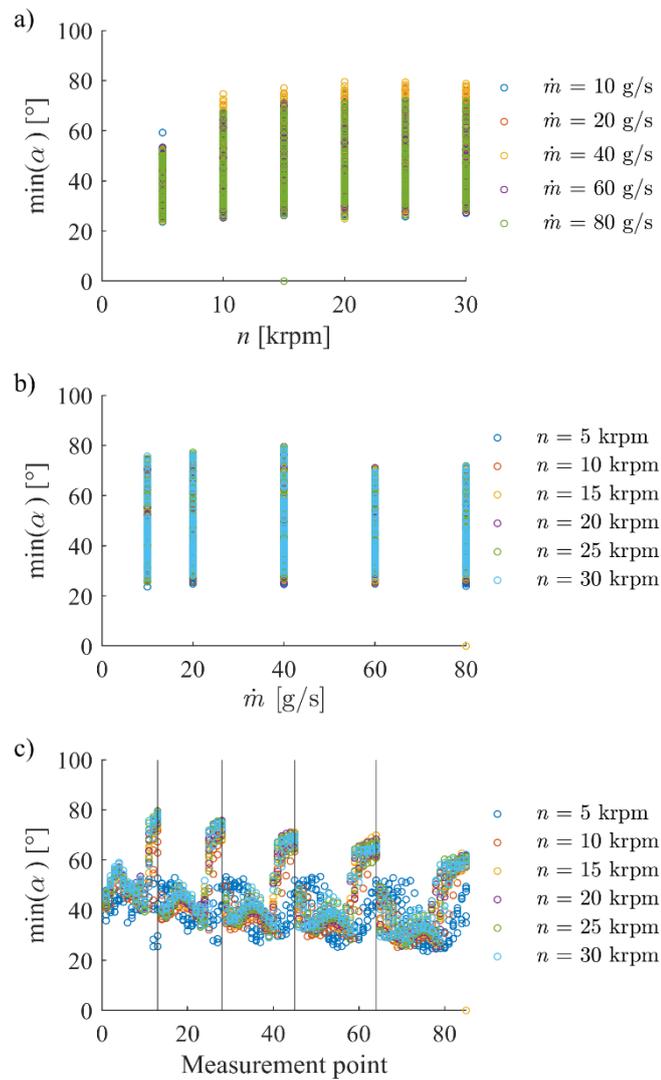

**Figure 10.** Minima of radial angles after filtering as a function of a) frequency, b) mass flow rate, c) measurement point.

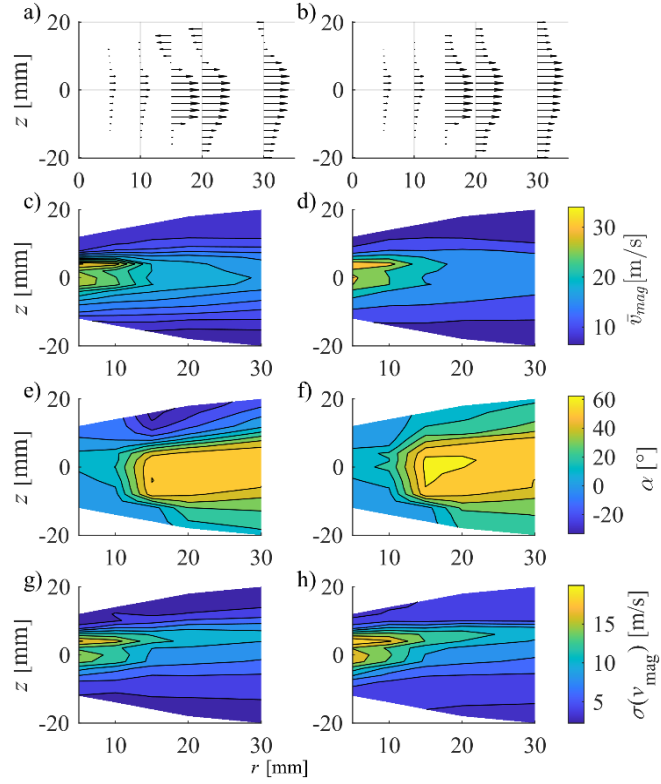

**Figure 11.** Results of filtering and velocity field on the *r–z* plane at *n* = 10 krpm and $\dot{m}$ = 40 g/s: first column before filtering, second column using $f_{v_r/v_t \vee v_r > 0}$, a) and b): velocity field, c) and d): mean total velocity, e) and f): mean radial angle, g) and h): standard deviation of total velocity.

Recirculation zones can be formed at the periphery, depending on the frequency, which causes negative radial droplet velocity. Figure 11 shows an example at *n* = 10 krpm and $\dot{m}$ = 40 g/s. Figure 11a shows the velocityfield , 11c shows the mean velocity magnitude, 11e shows the mean radial angle of the velocities, and 11g shows the standard deviation of the total velocity before filtering A strong recirculation zone appears at the *z* > 0 periphery which creates an asymmetrical flow field as α and $\bar{v}_{mag}$ shows. The presented behavior in Figs. 11 and 12 characterizes all the other data at various *n* values, except *n* = 5 krpm, which is detailed in Appendix B.

Figures 11b, 11d, 11f, and 11h shows the results after using $f_{v_r/v_t \vee v_r > 0}$. The recirculation zone disappeared, and a nearly symmetrical α and $\bar{v}_{mag}$ distributions are created in Figs. 11d and 11f.

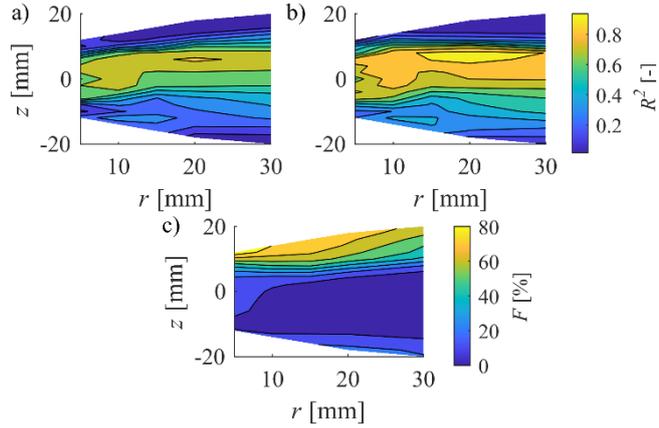

**Figure 12.** Results of filtering and velocity field on the *r*–*z* plane at *n* = 10 krpm and $\dot{m}$ = 40 g/s: a) $R_o^2$, b) $R_f^2$ using $f_{v_r/v_t \vee v_r > 0}$, c) filtered droplet ratio.

Figure 12a shows $R_o^2$, 12b $R_f^2$ using $f_{v_r/v_t \vee v_r > 0}$, and 12c shows *F*. *F* remains relatively low in the undisturbed flow, while in the periphery at *z* < 0 mm, there is a moderate increase in *F*. In the recirculation zone at *z* > 0 mm, at the spray periphery, *F* is significantly increased due to removing numerous negative radial velocity droplets. As Fig. 6b and 6d show, at the *z* < 0 mm periphery, a recirculation zone appears and results in lower *F*, so it has a weaker influence in the spray profile than at *z* > 0 mm spray periphery. Overall, based on Figs. 11 and 12, *F* is a good indicator of a recirculation zone.

Figure 12a shows that $R_o^2$ is asymmetrical and decreases towards the spray periphery. The maxima is close to the *z* = 0 and 4 mm lines where the two sets of orifices are located. Figure 12b shows $R_f^2$ after using $f_{v_r/v_t \vee v_r > 0}$, where $R^2$ is increased with global filtering in the *r*–*z* plane. As a result, $R_f^2$ can achive 90% in the central part of the spray, while it is low at the periphery, especially in the recirculation zone. Therefore, to achieve proper results in the recirculation zone and periphery, the $v_r > 0$ filter condition is necessary despite the high $F$ and moderate increase in $R_f^2$ values.

*3.3 Results of line fitting*

The slope, *A*, and the intercept, *B*, of the fitted line depend on the measurement settings. Figure 13 shows the *A–B* plane of all the 2550 datasets colored by different measurement parameters. In Figs. 13a and 13b, the original dataset shows no correlation between *A*, *B* and *n* or $\dot{m}$, such as the filtered dataset in Figs. 13c and 13d. Overall, using $f_{v_r/v_t \vee v_r > 0}$ slightly increases *A* but decreases *B*.

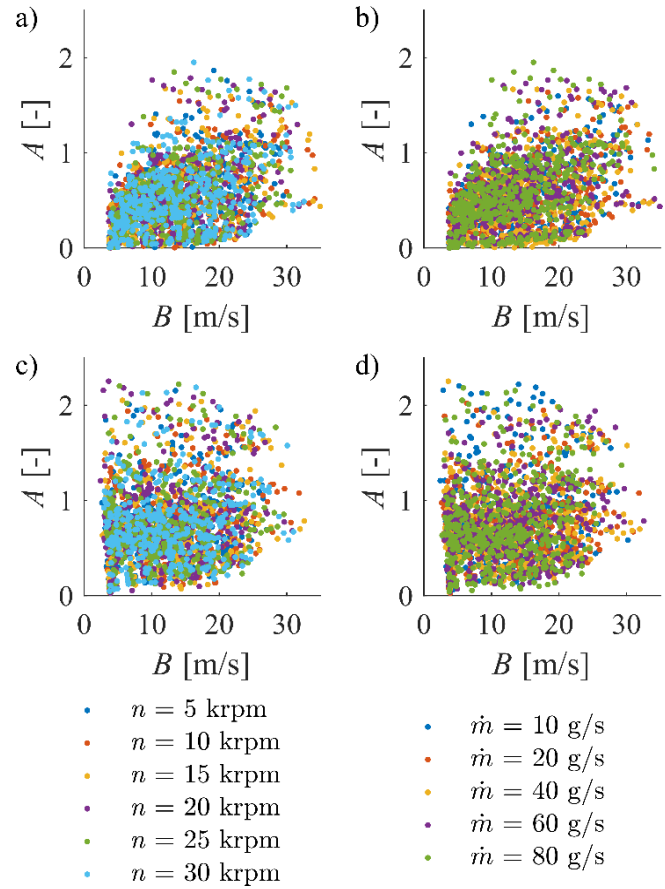

**Figure 13.** Relation of the slope and intercept of the fitted lines: a) and c) colored by $n$, b) and d) colored by $\dot{m}$, a) and b) without $f_{v_r/v_t \vee v_r > 0}$, c) and d) with the filter.

Figures 14a and 14b show the change of $A$, $\Delta A$, and Figs. 14c and 14d show the change of $B$, $\Delta B$, depending on the change of $R^2$, colored by $n$ and $\dot{m}$. By using the $f_{v_r/v_t \vee v_r > 0}$ filter, the majority of datasets result in increased $A$ with increased $R^2$; however, the minority show decreased $A$ with decreased $R^2$. In both cases, $B$ decreases, but with $R^2$, the change is more apparent. The effect of measurement setting is visible in the case of $A$, where increasing $\dot{m}$ causes a moderate change in $A$. Based on the results, there is no correlation with either $n$ or $\dot{m}$, the position in the spray has the highest influence.

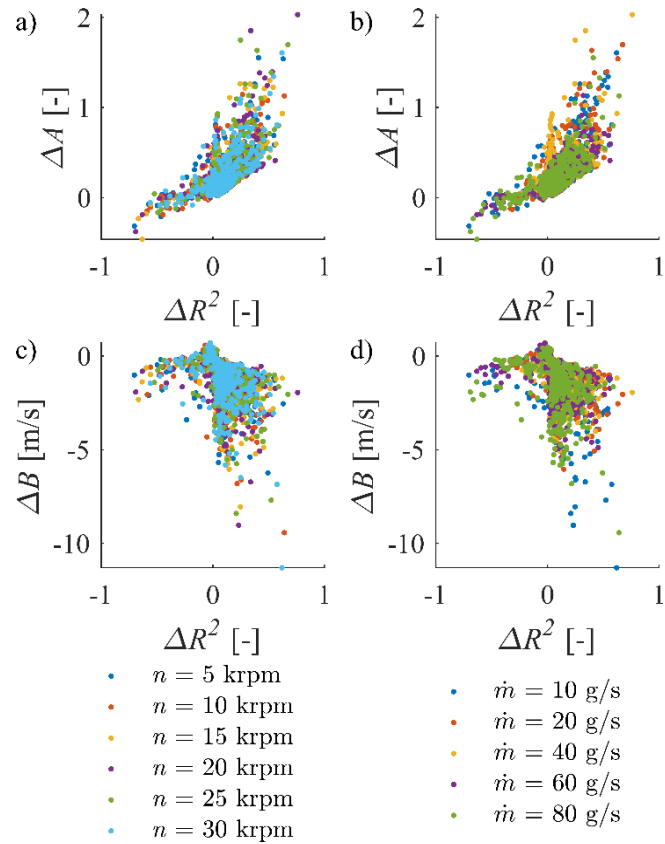

**Figure 14.** $\Delta A$ and $\Delta B$ as a function of $\Delta R^2$ due to the use of $f_{v_r/v_t \vee v_r > 0}$: a) and c) colored by $n$, b) and d) colored by $\dot{m}$.

Figure 15 shows *A* and *B* on the *r*–*z* measurement plane at *n* = 10 krpm and $\dot{m}$ = 40 g/s both in the case of the original and the filtered datasets. Figures 15a and 15c show the results of the original data, while Figs. 15b and 15d show the results of the filtered data. By comparing Figs. 15a to 15b and 15c to 15d, the results show the same conclusion as Fig. 13. Namely, filtering increases *A* while decreasing *B*. With increasing *r* and increasing *z* in absolute value, both *A* and *B* decrease.

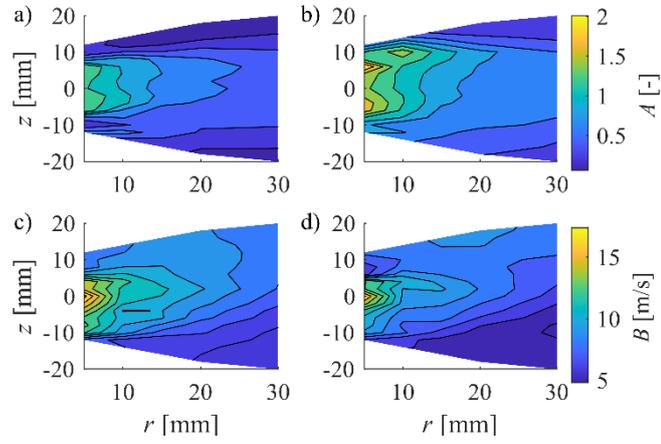

**Figure 15.** Result of *A* and *B* on the *r*–*z* plane at *n* = 10 krpm and $\dot{m}$ = 40 g/s: the first row shows the results of *A*, while the second row shows the results of *B*. The original data is shown in a) and c), while data with a filter is shown in b) and d).

The *r*–*z* profiles are asymmetrical due to the two sets of orifices with different diameters, positioned at *z* = 0 mm and *z* = 4 mm. Therefore, it is necessary to investigate the effect of *r* and *z*. Figure 16 shows the mean of *A* for all $\dot{m}$ by *n*, depending on *r* along different *z* lines. By increasing *r*, *A* decreases except in Fig. 16d. Along the *z* = 8 mm line, the decreasing trend of *A* is distorted due to recirculation. In Figs. 16b and 16d, near the central region of the measurement plane, *A* increases slightly with increasing *n*. Figure 17 shows the same trends of *A* but at different $\dot{m}$.

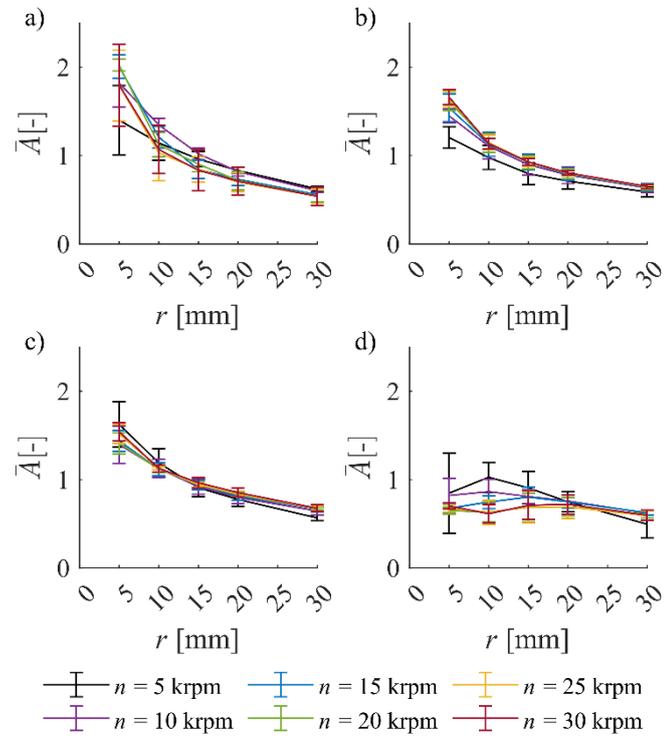

**Figure 16.** Mean of *A* of all $\dot{m}$ by *n* as a function of *r*: a) *z* = -8 mm, b) *z* = 0 mm, c) *z* = 4 mm and d) *z* = 8 mm. The errorbar shows the standard deviation of *A* at different $\dot{m}$ at a given *n*.

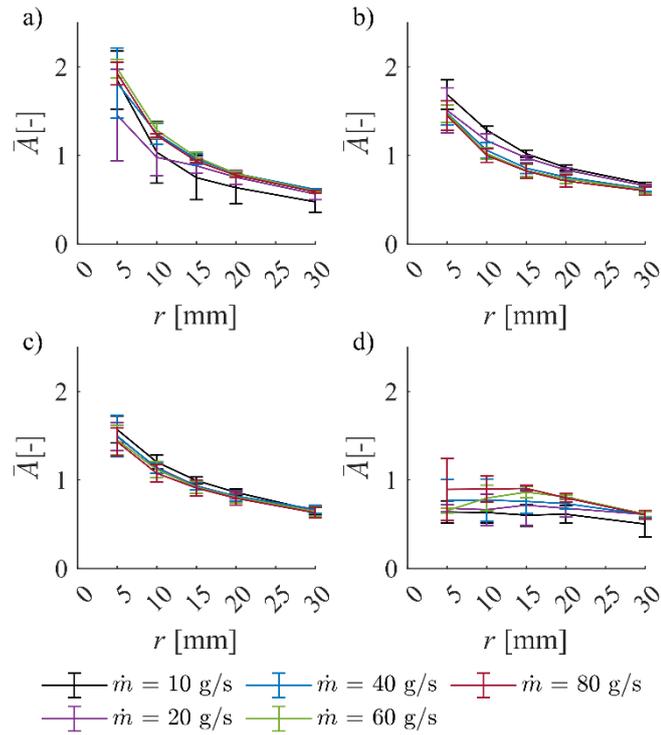

**Figure 17.** Mean of *A* of all by $\dot{m}$ as a function of *r*: a) $z = -8$ mm, b) $z = 0$ mm, c) $z = 4$ mm and d) $z = 8$ mm. The errorbar shows the standard deviation of *A* at different *n* at a given $\dot{m}$.

Figure 18 shows the mean of *B* at all $\dot{m}$ by *n*, depending on *x*, along different *y* lines. B decreases with increasing *r*, which is more significant towards $z > 0$ mm lines. The effect of *n* is more significant than in the case of *A*. Figure 19 shows the relation between the mean of *B* of all *n* and *r* but at different $\dot{m}$. The trends are similar compared to Fig. 18, but the effect of $\dot{m}$ is less significant than that of *n*.

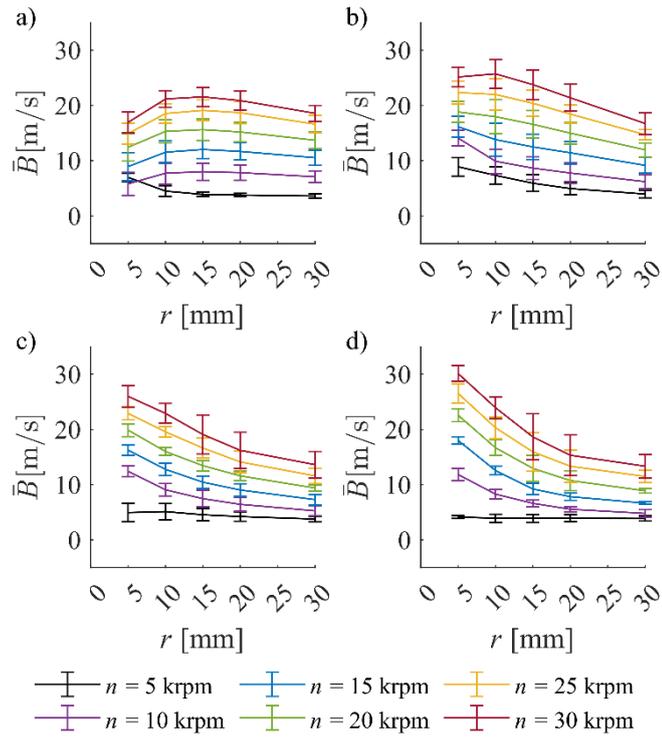

**Figure 18.** Mean of $B$ of all $\dot{m}$ by $n$ as a function of $r$: a) $z = -8$ mm, b) $z = 0$ mm, c) $z = 4$ mm and d) $z = 8$ mm. The errorbar shows the standard deviation of $B$ at different $\dot{m}$ at a given $n$.

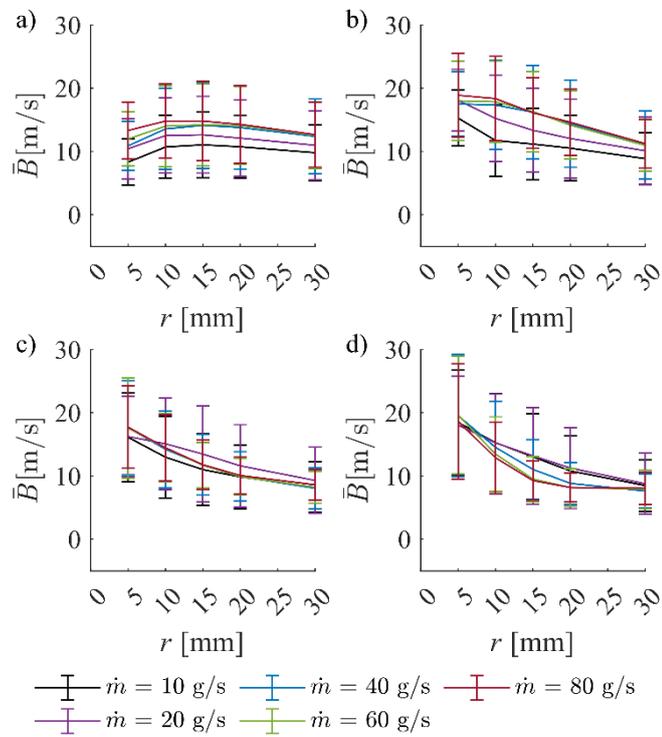

**Figure 19.** Mean of *B* of all by $\dot{m}$ as a function of *r*: a) *z* = -8 mm, b) *z* = 0 mm, c) *z* = 4 mm and d) *z* = 8 mm. The errorbar shows the standard deviation of *B* at different *n* at a given $\dot{m}$.

## 5. Conclusions

Rotary atomizers generate a multiphase flow, where droplets possess significant radial and tangential velocities. The present work showed a comprehensive flow field evaluation by exploiting the relationship between the two velocity components. The measurements were performed by a two-component Phase Doppler Anemometer, using a rotary atomizer with two hole sets; the atomized fluid was water. Based on the results, the following conclusions are made:

The measurement requires a special setup to keep the optical instruments clean, inevitably biasing the measurement results. Therefore, the first part of this paper evaluated ten different data filters to remove biased droplets. The best one was a combination of the radial-to-tangential velocity ratio and positive radial velocity, $f_{v_r/v_t \vee v_r > 0}$.

Most data are concentrated around a line in a $v_r$–$v_t$ plot at each measurement point with varying slope and intercept values. The applied filter efficiently removed droplets from alternative sources, which often exerted a high momentum on the fitted line. Consequently, the $R^2$ of the fits generally increased.

The physical soundness of the applied filter is evaluated via its effect on the angle of the velocity components as a function of both $v_r$ and $v_t$. Ultimately, the recirculation zones were clearly identified, where data sets became irregular.

Finally, the slope and intercept of the fitted lines were presented. The mean of the former parameter collapsed into a single curve as a function of the radial distance from the orifice. In contrast, the latter parameter showed a more condition-dependent trend.

**Acknowledgments**

The research reported in this paper was supported by the National Research, Development and Innovation Fund of Hungary, project №. OTKA-FK 137758, and TKP2021 Grant №. BME-NVA-02, ÚNKP-23-5-BME-449, and ÚNKP-23-3-II-BME-405 New National Excellence Program of the Ministry for Culture and Innovation from the source of the National Research, Development and Innovation Fund. The authors acknowledge the financial support from project no. 21-45227L funded by the Czech Science Foundation and project no. FSI-S-23-8192 founded by Brno University of Technology.

**Declarations**

- Conflict of Interest: Declaration of interests

    ☐ The authors declare that they have no known competing financial interests or personal relationships that could have appeared to influence the work reported in this paper.

    ☒ The authors declare the following financial interests/personal relationships which may be considered as potential competing interests:

    Erika Rácz reports financial support was provided by Ministry for Innovation and Technology Hungary. Viktor Józsa reports financial support was provided by Ministry for Innovation and Technology Hungary. Jan Jedelský reports financial support was provided by Czech Science Foundation. Milan Malý reports financial support was provided by Czech Science Foundation. Ondřej Cejpek reports financial support was provided by Czech Science Foundation. If there are other authors, they declare that they have no known competing financial interests or personal relationships that could have appeared to influence the work reported in this paper. No ethical issue was emerged during this research project.

- Author contributions

    **Erika Rácz**: Conceptualization, Methodology, Software, Validation, Formal analysis, Data curation, Writing - Original Draft, Writing - review & editing, Visualization

    **Milan Malý**: Validation, Investigation, Resources, Supervision

    **Ondřej Cejpek**: Validation, Investigation, Resources

    **Jan Jedelský**: Resources, Supervision, Project administration, Funding acquisition.

    **Viktor Józsa**: Conceptualization, Methodology, Validation, Writing - Original Draft, Writing - review & editing, Supervision, Project administration, Funding acquisition.

- Data availability

    The data that support the findings of this study are available from the corresponding author upon reasonable request.

**Appendix A**

Appendix A contains the radial and tangential velocity distribution, similar to Fig. 5, to show the effect of alternative, undiscussed filters. Figure A.1 shows the result of the $f_D$ filter. The outlying droplets are rarely the high $v_r$ and low $v_t$ ones, regardless of their obvious alternative trend. Figures A.1a and A.1c represent the central part of the spray, where the filter removes the large droplets close to the fitted line. This result is undesired because these droplets reflect the underlying physical phenomena correctly. Figures A.1b and A.1d representing the periphery of the spray, where the filter removes some droplets in the center, meaning no significant contribution to data quality.

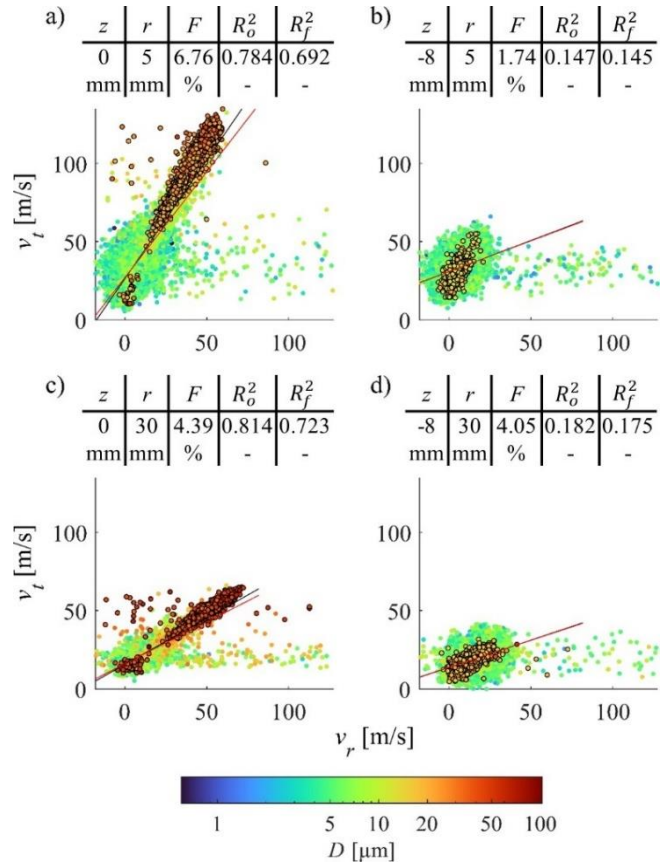

**Figure A.1.** Radial and tangential velocity distribution of the spray, colored by the droplet diameter at $n = 30$ krpm and $\dot{m} = 10$ g/s, using the $f_D$ filter. The black line is fitted to the original data set, while the red line is fitted to the filtered data set; the discarded droplets are marked with black circles.

Figure A.2 shows the effect of the $f_{v_t}$ filter. It removes the outlying, high- and low-tangential velocity droplets. It leads to physically inadequate results again. The filter leaves the outlying high $v_r$ low $v_t$ droplets intact while removing a significant amount of droplets near the fitted line. Therefore, $R^2$ decreases as the slope of the fitted line.

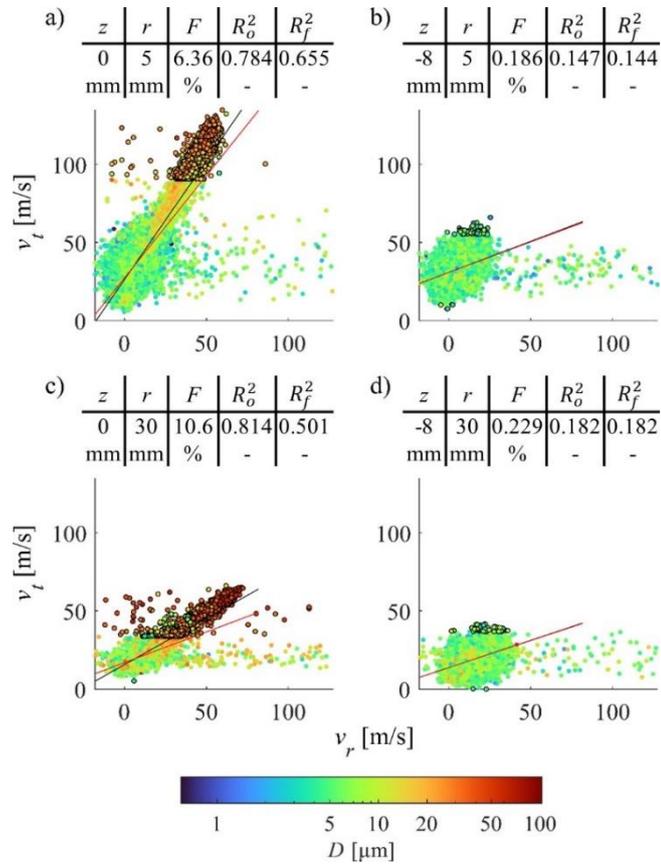

**Figure A.2.** Radial and tangential velocity distribution of the spray, colored by the droplet diameter at $n = 30$ krpm and $\dot{m} = 10$ g/s, using the $f_{v_t}$ filter. The black line is fitted to the original data set, while the red line is fitted to the filtered data set; the discarded droplets are marked with black circles.

Figure A.3 shows the results of the $f_{v_r}$ filter. It is superior to the $f_{v_t}$ filter as droplets with high $v_r$ and low $v_t$ are removed. Nevertheless, droplets with high $v_r$ and high $v_t$ are unnecessarily removed. Figures A.3b and A.3c represent the periphery of the spray. Due to the oval shape of the distribution and the removal of negative-radial velocity droplets, moderate improvements are realized in the line fitting. Overall, using $f_{v_r}$ leads to the same conclusion as in the case of using $f_{v_t}$ with marginal improvements.

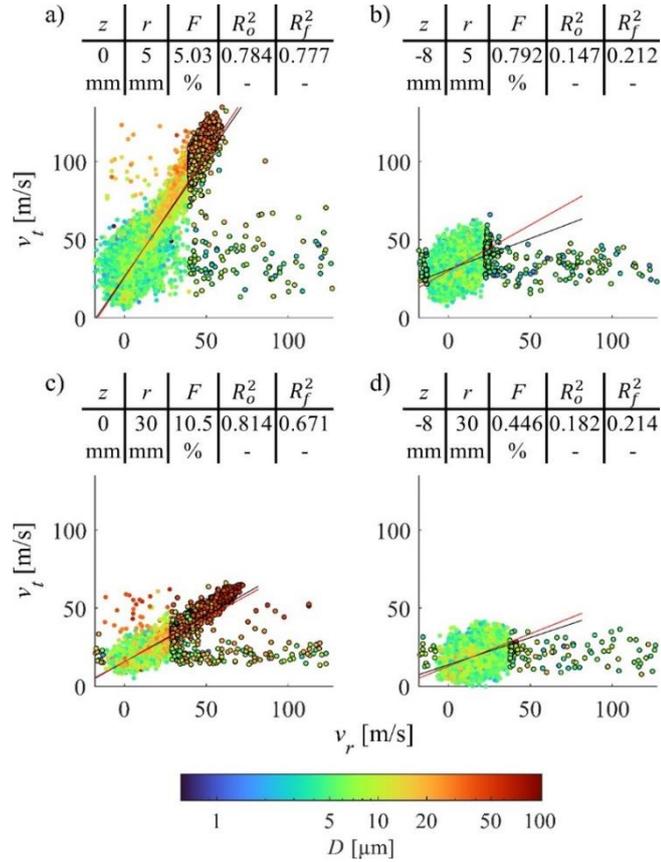

**Figure A.3.** Radial and tangential velocity distribution of the spray, colored by the droplet diameter at $n = 30$ krpm and $\dot{m} = 10$ g/s, using the $f_{v_r}$ filter. The black line is fitted to the original data set, while the red line is fitted to the filtered data set; the discarded droplets are marked with black circles.

Based on the results of $f_{v_r}$ and $f_{v_t}$, it is obvious that a filter with the combination of both velocity components is required to follow the theoretical behavior of the rotary spray. A further possible option is outlier filtering by $f_{v_{mag}}$; the results are shown in Fig. A.4. This filter leads to the same results, such as $f_{v_r}$ by the overfiltering of the droplets near the fitted line. However, many high $v_r$ and low $v_t$ droplets are removed.

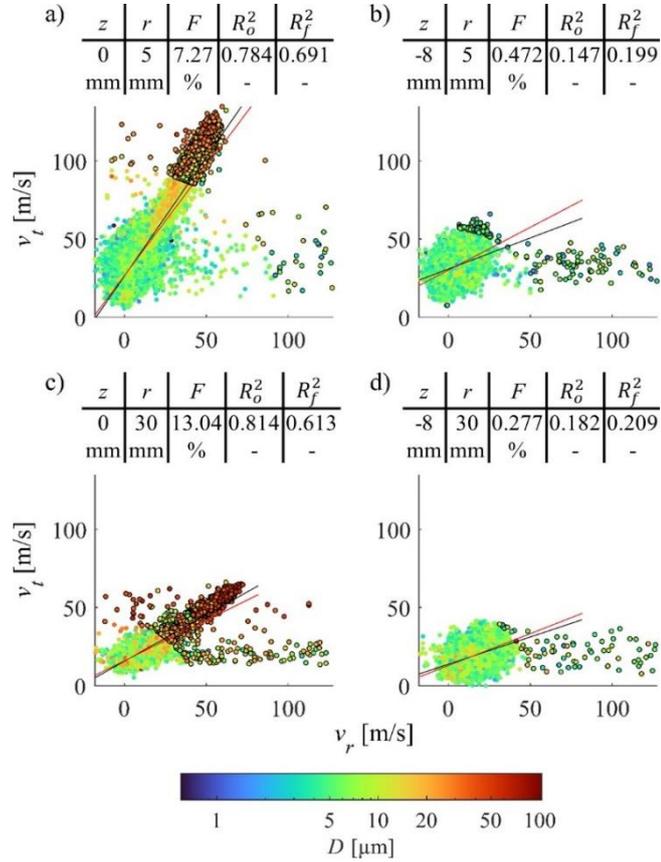

**Figure A.4.** Radial and tangential velocity distribution of the spray, colored by the droplet diameter at $n = 30$ krpm and $\dot{m} = 10$ g/s, using the $f_{v_{mag}}$ filter. The black line is fitted to the original data set, while the red line is fitted to the filtered data set; the discarded droplets are marked with black circles.

For removing droplets with high $v_r$ and $v_t$, another option is to apply $f_{v_r \vee v_t}$, the combination of $f_{v_r}$ and $f_{v_t}$; the results are shown in Fig. A.5. The improvement was marginal, as shown in Table II. Even though droplets with high $v_r$ or $v_t$ were filtered, significant data points were removed from the high $v_r$ and $v_t$ droplets, meaning a notable drawback to keeping the physically significant characteristics.

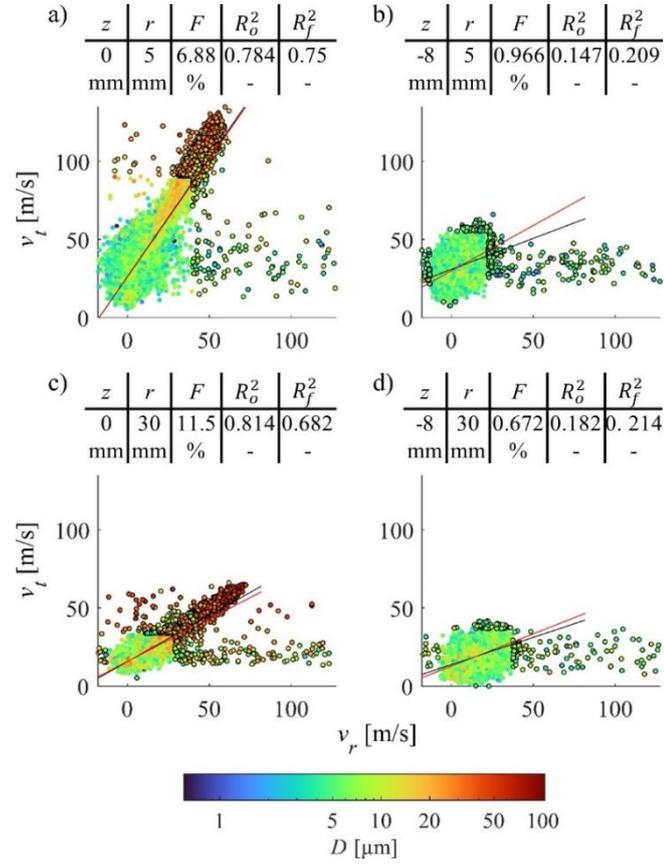

**Figure A.5.** Radial and tangential velocity distribution of the spray, colored by the droplet diameter at $n = 30$ krpm and $\dot{m} = 10$ g/s, using the $f_{v_r \vee v_t}$ filter. The black line is fitted to the original data set, while the red line is fitted to the filtered data set; the discarded droplets are marked with black circles.

Figure A.6. shows the results of $f_{v_r > 0}$, which removes the negative radial velocity droplets appearing in the data due to the recirculation in the closed measurement chamber. $\Delta R^2$ does not show any significant change, regardless of whether $F$ is notable. Nevertheless, using $f_{v_r > 0}$ is necessary to remove droplets, which do not belong to the fresh spray.

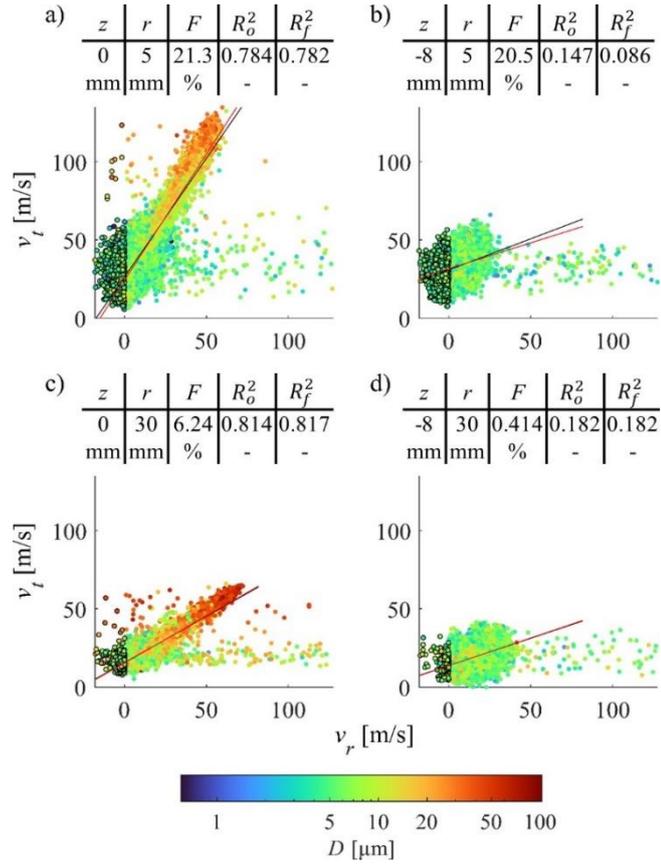

**Figure A.6.** Radial and tangential velocity distribution of the spray, colored by the droplet diameter at $n = 30$ krpm and $\dot{m} = 10$ g/s, using the $f_{v_r > 0}$ filter. The black line is fitted to the original data set, while the red line is fitted to the filtered data set; the discarded droplets are marked with black circles.

The outlying droplets have high $v_r$ and low $v_t$ or high $v_t$ and low $v_r$, and they are far from the fitted line. The velocity ratio of these droplets differs from the ones close to the fitted line; therefore, it is possible to identify them by checking the ratio of the velocity components. The $f_{v_r/v_t}$ makes outlier filtering by the ratio of the radial and tangential velocity components, and the results are shown in Fig. A.7. Due to the negative $v_r$ values and the right-skewed distribution, the $v_r/v_t$ filter eliminates both negative and positive outlier values. This filter avoids over-filtering close to the fitted line and increases $R^2$ while keeping $F$ low. If this is critical $f_{v_r/v_t}$ is the best filter for data processing.

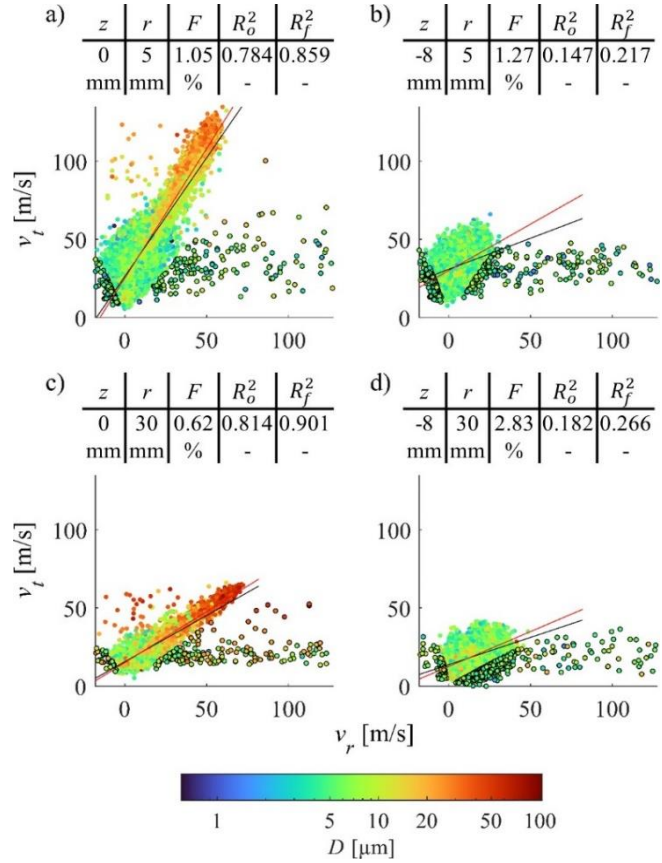

**Figure A.7.** Radial and tangential velocity distribution of the spray, colored by the droplet diameter at $n = 30$ krpm and $\dot{m} = 10$ g/s, using the $f_{v_r/v_t}$ filter. The black line is fitted to the original data set, while the red line is fitted to the filtered data set; the discarded droplets are marked with black circles.

Figure A.8. shows the results of the $f_{v_t/v_r}$ filter, which is the reciprocal of the previous one. It is unsuitable for filtering outliers because it leaves droplets with high $v_r$ and low $v_t$ intact and removes droplets only whose $v_r/v_t$ ratio significantly differs from the slope of the fitted line. The filtered droplets have relatively high $v_t$ and low $v_r$ in both negative and positive ranges. Therefore, it removes droplets with $v_r$ close to zero and cuts the data incorrectly.

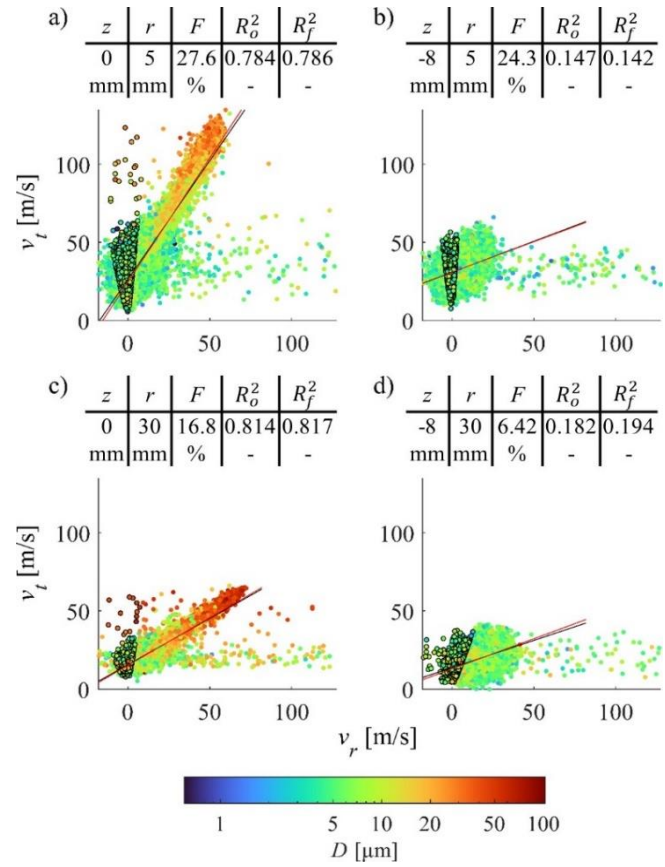

**Figure A.8.** Radial and tangential velocity distribution of the spray, colored by the droplet diameter at $n = 30$ krpm and $\dot{m} = 10$ g/s, using the $f_{v_t/v_r}$ filter. The black line is fitted to the original data set, while the red line is fitted to the filtered data set; the discarded droplets are marked with black circles.

**Appendix B**

Appendix B shows the results of the line fitting and the velocity field at $n = 5$ krpm data. This value is highlighted due to the different behavior compared to the other $n$ values. Figure B.1 shows the emerging flow field at $n = 5$ krpm and $\dot{m} = 40$ g/s. Figure B.1a, compared to Fig. 10a, shows a different velocity profile; the maxima are shifted towards the $z > 0$ mm region. Near the nozzle, around $z = 0$ mm, a small inner recirculation zone is present, which is absent at higher $n$, and there is no recirculation zone at the $z > 0$ periphery of the spray. As Figs. B.1d and B.1f show the data due to recirculation is removed, but the symmetry of the flow field is not achived in $\alpha$, but in $\bar{v}_{mag}$.

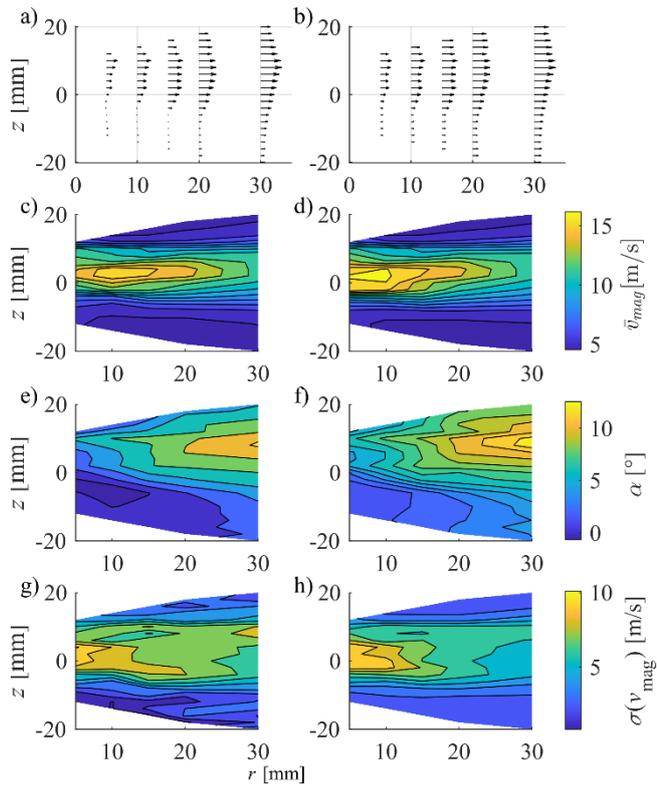

**Figure B.1.** Results of filtering and velocity field on the $r$–$z$ plane at $n = 5$ krpm and $\dot{m} = 40$ g/s: first column before filtering, second column using $f_{v_r/v_t \vee v_r > 0}$, a) and b): velocity field, c) and d): mean total velocity, e) and f): mean radial angle, g) and h): standard deviation of total velocity.

Figures B.1a and B.1b show increasing $R^2$ at the central region of the spray and higher $r$ values, especially in the recirculation zone. However, as Fig B.1c shows, the strong correlation between $F$ and the appearance of the recirculation zone is present at $n = 5$ krpm, too.

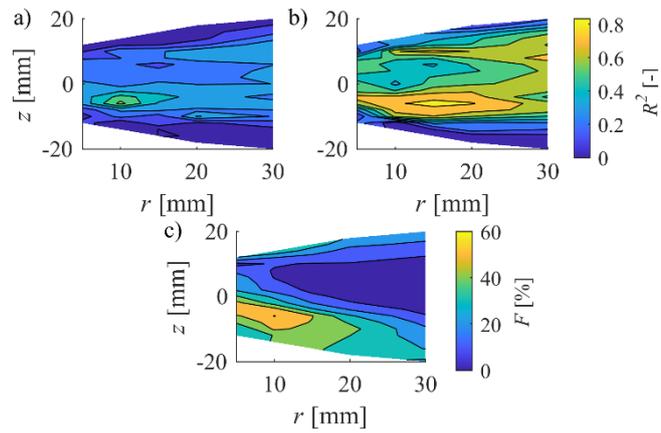

**Figure B.2.** Results of filtering and velocity field on the $r$–$z$ plane at $n = 5$ krpm and $\dot{m} = 40$ g/s: a) $R_o^2$, b) $R_f^2$ using $f_{v_r/v_t \vee v_r > 0}$, c) filtered droplet ratio.